\documentclass[aps, twocolumn,pra,superscriptaddress]{revtex4-1}
\usepackage{graphicx}
\usepackage[utf8]{inputenc}
\begin{document}

\title{Optomechanical-like coupling between superconducting resonators} 

\author{J.R. Johansson}
\email{robert@riken.jp}
\affiliation{iTHES Research Group, RIKEN, Wako-shi, Saitama, 351-0198 Japan}
\author{G. Johansson}
\affiliation{Microtechnology and Nanoscience, MC2, Chalmers University of Technology, SE-412 96 G{\"o}teborg, Sweden}
\author{Franco Nori}
\affiliation{CEMS, RIKEN, Wako-shi, Saitama, 351-0198 Japan}
\affiliation{Department of Physics, University of Michigan, Ann Arbor, Michigan 48109-1040 USA}

\begin{abstract}
We propose and analyze a circuit that implements a nonlinear coupling between
two superconducting microwave resonators. The resonators are coupled through a superconducting
quantum interference device (SQUID) that terminates one of the resonators. This produces a nonlinear
interaction on the standard optomechanical form, where the
quadrature of one resonator couples to the photon number of the other resonator. The circuit therefore
allows for all-electrical realizations of analogs to optomechanical systems, 
with coupling that can be both strong and tunable.
We estimate the coupling strengths that should be attainable with the proposed device,
and we find that the device is a promising candidate for 
realizing the single-photon strong-coupling regime.
As a potential application, we discuss implementations of networks of nonlinearly-coupled microwave resonators,
which could be used in microwave-photon based quantum simulation.
\end{abstract}

\date{\today}
% PACS numbers:
% 42.50.Wk : mechanical effects on atoms and molecules
% 42.50.Pq : Cavity quantum electrodynamics
% 85.25.Cp : Josephson devices
\maketitle

\section{Introduction}\label{sec:intro}

Superconducting microwave resonators have emerged as one of the key components in quantum electronics \cite{you:2005,clarke:2008,you:2011,devoret:2013} in recent years.
In a parallel development, the field of quantum optomechanics \cite{poot:2012,meystre:2013,aspelmeyer:2013} have seen equally impressive progress,
with recent accomplishments including sideband cooling of mechanical resonators to their ground state \cite{teufel:2011:2, chan:2011}, normal-mode splitting \cite{groblacher:2009,teufel:2011:1,verhagen:2012}, generation of
nonclassical states of light \cite{brooks:2012,safavi-naeini:2013}, near quantum-limited detection \cite{regal:2008,teufel:2009,hertzberg:2010}, and state transfer \cite{palomaki:2013, palomaki:2013:2}. 
In several of these recent works \cite{teufel:2011:1,hertzberg:2010, teufel:2011:2, regal:2008,teufel:2009,palomaki:2013, palomaki:2013:2}, microwave resonators, rather than optical cavities, were coupled to the mechanical systems.
Meanwhile, in electrical systems, superconducting microwave resonators have been used as quantum buses to couple superconducting qubits in a variety of architectures \cite{you:2003,blais:2004,schoelkopf:2008}, for readout and control of superconducting qubits \cite{duty:2005,wallraff:2005,siddiqi:2006,lupascu:2006}, for characterization of quantum dots \cite{delbecq:2011,chorley:2012,frey:2012,colless:2013}, and for interfacing different types of quantum systems in hybrid circuits \cite{xiang:2013}. 

\begin{figure}[b]
\begin{center}
\includegraphics[width=8.5cm]{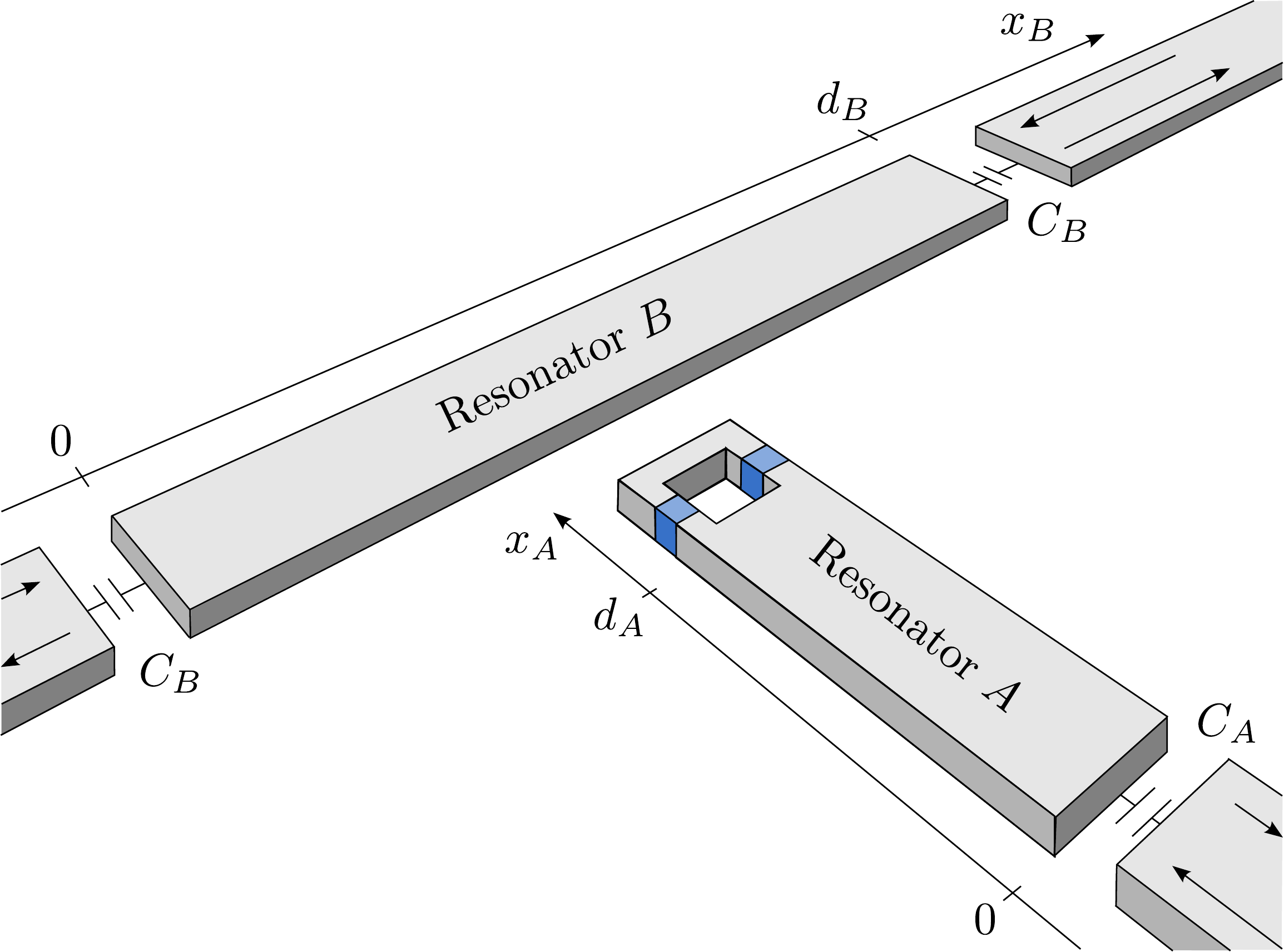}
\caption{(color online) A schematic illustration of the system, which consists of two superconducting transmission-line resonators $A$ and $B$. The resonators are coupled to each other though the SQUID terminating resonator $A$. The coupling mechanism is the following: part of the magnetic field generated by the signal in resonator $B$ threads the SQUID loop that terminates resonator $A$, changing the phase across the SQUID. This phase determines the boundary condition for resonator $A$. The result is an interaction where the amplitude of resonator $B$ couples to the photon number of resonator $A$.}
\label{fig:io-model}
\end{center}
\end{figure}

Coupled microwave resonators has also been studied extensively, both theoretically \cite{zhou:2008,helmer:2009,schmidt:2010} and experimentally \cite{underwood:2012}. However, in these circuits, the resonators are typically coupled linearly to each other or to other quantum systems, through the amplitude of the resonator's electric or magnetic field. Here we investigate a nonlinear coupling between two microwave resonators, where, in a certain regime, the field amplitude in one resonator couples to the photon number in the other resonator. This is exactly the type of interaction encountered in optomechanical systems \cite{poot:2012,meystre:2013,aspelmeyer:2013}, making it possible to implement analogs of optomechanical systems in all-electrical circuits. In such analogs, the mechanical component is replaced by an electrical resonator, but without losing the interesting nonlinear coupling that is characteristic for optomechanical systems. Moreover, using this type of device, it appears possible to reach the single-photon strong-coupling regime. This regime has recently received considerable attention, and a number of interesting phenomena has been theoretically predicted, including photon blockade effects \cite{rabl:2011}, multiple cooling resonances \cite{nunnenkamp:2011, nunnenkamp:2012}, and the generation of nonclassical states \cite{bose:1997,qian:2012,kronwald:2013,nation:2013}.

The physical realization of this nonlinear coupling uses a SQUID embedded in one of the resonators. The magnetic flux that threads the SQUID-loop can modify the properties of the resonator, such as its resonance frequency \cite{wallquist:2006}. Superconducting microwave resonators with embedded SQUIDs have been used to implement frequency-tunable resonators \cite{laloy:2008,sandberg:2008,yamamoto:2008} with tunable boundary conditions and tunable index of refraction. With parametrically modulated applied magnetic flux, i.e., with classical driving fields, these types of devices have been used to implement parametric amplifiers \cite{yamamoto:2008,castellanos:2008,wilson:2010} and nonadiabatic quantum phenomena such as the dynamical Casimir effect \cite{johansson:2009, johansson:2010, johansson:2013, wilson:2011, lahteenmaki:2013}. See, e.g., Ref.~\cite{nation:2012} for a recent review. 

Here we are interested in the case when the applied magnetic flux through the SQUID is due to the quantum field of another superconducting resonator, i.e., a quantized drive field. Also, we consider the situation where the modulated resonator adiabatically adjust to the changes imposed by the magnetic flux though the embedded SQUID. Under these conditions we can formulate an effective Hamiltonian that describes the dynamics of the system. We show that this effective Hamiltonian is on the standard optomechanical form.

The remaining part of this paper is organized as follows: In Sec.~\ref{sec:circuitmodel} we introduce
the device and setup a model for it. Here we use the Lagrangian formalism to
model a lump-element representation of the circuit to obtain the boundary conditions
and finding the adiabatic mode functions for the resonators. In
Sec.~\ref{sec:effective-hamiltonian} we use the derived mode functions to formulate
an effective Hamiltonian for the system, which is shown to be on the optomechanical
form in Sec.~\ref{sec:optomech-hamiltonian}. In Sec.~\ref{sec:coupling-strength} we
analyze possible coupling designs and evaluate the corresponding coupling strengths.
In Sec.~\ref{sec:arrays} we discuss possible circuit layouts for realizing arrays
of nonlinearly coupled resonators. Finally, we summarize our results in
Sec.~\ref{sec:conclusions}.

\section{The device and its circuit model} \label{sec:circuitmodel}

The type of device we investigate here is shown in one possible configuration in Fig.~\ref{fig:io-model}. Alternative configurations could also be used, with for example a SQUID located in the middle of resonator $A$, or with resonator $A$ made of an array of SQUIDs. The main properties of the system would remain unchanged.

Here we focus on a quantum mechanical analysis of the device in Fig.~\ref{fig:io-model}. The flux through the SQUID can in a certain regime be seen as modulating the effective length of resonator $A$, or equivalently, its fundamental resonance frequency $\omega_A$. The flux through the SQUID is partly due to the magnetic field generated by resonator $B$. We therefore expect an interaction on the form $a^\dagger a(b+b^\dagger)$, where $a$ and $b$ are the annihilation operators for resonator $A$ and $B$, respectively. In the following we derive this result from a detailed quantum network analysis \cite{yurke:1984, devoret:1995} of the circuit.

%-------------------------------------------------------------------------------
% Lagrangian
%
\subsection{Circuit Lagrangian}\label{sec:lagrangian}

\begin{figure}[t]
\begin{center}
\includegraphics[width=8.5cm]{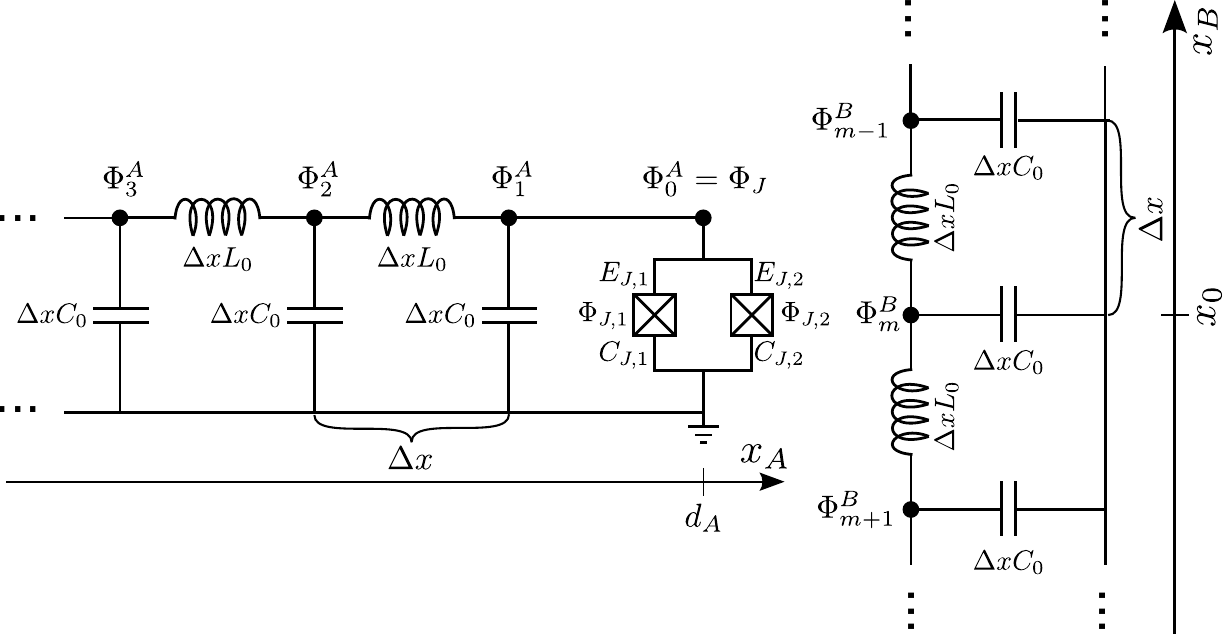}
\caption{A circuit diagram of the coupler part of the circuit in Fig.~\ref{fig:io-model},
where resonator B meets the SQUID embedded in resonator A.}
\label{fig:squid-coupler-circuit-diagram}
\end{center}
\end{figure}

We model the electrical circuit in Fig.~\ref{fig:io-model} by decomposing it in lumped-circuit elements,
as shown, for the most relevant part of the circuit, in Fig.~\ref{fig:squid-coupler-circuit-diagram}.
As generalized coordinates we use the magnetic node fluxes $\Phi_n^{\alpha}$ (where $\alpha =$ A, B),
which are related to the node voltages $V_n^\alpha$ as $\Phi_n^\alpha = \int^t dt' V_n^\alpha(t')$,
and to the gauge-invariant superconducting node phases $\varphi_n^\alpha = 2\pi\Phi_n^\alpha/\Phi_0$.
In terms of these coordinates, the Lagrangian of the circuit can be written as
\begin{equation}
\mathcal{L} = \mathcal{L}_A + \mathcal{L}_B + \mathcal{L}_{S}
\end{equation}
where
\begin{eqnarray}
\mathcal{L}_\alpha &=& 
\frac{1}{2}\sum_{n=1}^{N_\alpha} \left(\Delta x C_0^\alpha (\dot{\Phi}_n^\alpha)^2 - \frac{(\Phi_{n+1}^\alpha - \Phi_n^\alpha)^2}{\Delta x L_0^\alpha}\right), \\
\mathcal{L}_{\rm S} &=& 
\frac{1}{2}C_J \left(\dot\Phi_J\right)^2 + E_J(\Phi_{\rm ext}) \cos\left(2\pi \frac{\Phi_J}{\Phi_0}\right).
\end{eqnarray}
Here we have assumed that the SQUID is symmetric ($C_{J,i}=C_{J}$ and $E_{J,i}=E_J$) and we have written its Lagrangian $\mathcal{L}_{\rm S}$ on the form of an effective Josephson junction with Josephson energy
\begin{equation}
\label{eq:SQUID_E_J}
E_J(\Phi_{\rm ext}) = 2E_J\left|\cos\left(\pi\frac{\Phi_{\rm ext}}{\Phi_0}\right)\right|.
\end{equation}
In the following we also assume that the two transmission lines are uniform, with $C_0^\alpha = C_0$ and $L_0^\alpha = L_0$.

From the circuit Lagrangian, we obtain equations of motion for flux nodes $\Phi_n^{\alpha}$. 
In the continuum limit, $\Delta x \rightarrow 0$, the resulting flux fields $\Phi_A(x,t)$ and $\Phi_B(x,t)$ are found
to obey the one-dimensional massless Klein-Gordon wave equation, which has a continuum
of independent plane-wave solutions propagating in the positive and negative direction, 
respectively. We can therefore write the quantum mechanical representation of the flux field as

\begin{eqnarray}
\label{eq:quantum_field}
\Phi(x) = \sqrt{\frac{Z_0\hbar}{4\pi}}\int \frac{\mathrm{d}\omega}{\sqrt{|\omega|}}
(a_R(\omega) e^{-i(-k_\omega x + \omega t)} \nonumber\\
+ a_L(\omega) e^{-i(k_\omega x + \omega t)} + \text{h.c.}),
\end{eqnarray}
where $a_L(\omega)$ and $a_R(\omega)$ are annihilation operators for the fields propagating
in the negative and positive $x$-direction, respectively, satisfying the commutation relation
$[a(\omega), a^\dagger(\omega')] = \delta(\omega - \omega')$. $Z_0 = \sqrt{L_0/C_0}$
is the characteristic impedance, $k_\omega = \omega/v$ is the wave number,
and $v = 1/\sqrt{L_0C_0}$ the propagation speed of the signal in the transmission line.
At the boundaries, i.e., at $x=0$ and $x=d_A$ for resonator $A$, and at $x=0$ and $x=d_B$ for resonator
$B$ (see Fig.~\ref{fig:io-model}), the equations of motion define the boundary conditions for the continuum fields $\Phi_A(x)$
and $\Phi_B(x)$. These boundary conditions can be used to derive the mode functions for the resonators.

%-------------------------------------------------------------------------------
% Boundary conditions
%
\subsection{Boundary conditions}\label{sec:bc}

In this section we write down the boundary conditions for the two sides of the 
two resonators. Here we assume that the two resonators have well-defined resonance
frequencies, i.e., their quality factors are high, and the capacitive coupling to
the external transmission lines shown in Fig.~\ref{fig:io-model} can be neglected
(that is, we consider the limit $C_{A,B} \rightarrow 0$). The boundary conditions
therefore correspond to that of an open-ended resonator.

\subsubsection{Resonator A}\label{sec:bc-res-A}

In the limit $C_A\rightarrow0$, resonator $A$ is open-ended at
$x = 0$ (see Fig.~\ref{fig:io-model}), and the corresponding boundary condition is
\begin{eqnarray}
\label{eq:bc_res_A_left}
\partial_x \Phi_A(0, t) &=& 0.
\end{eqnarray}
At the end terminated by the SQUID, $x = d_A$, the boundary condition \cite{wallquist:2006,johansson:2010}
can be written as
\begin{eqnarray}
\label{eq:bc_res_A_squid}
C_J \partial_{tt} \Phi_A(d_A, t) &+& \left(\frac{2\pi}{\Phi_0}\right)^2 E_J(\Phi_{\rm ext})\Phi_A(d_A, t) \nonumber\\
&+& \frac{1}{L_0}\partial_x \Phi_A(d_A, t) = 0.
\end{eqnarray}
If the plasma frequency of the SQUID is large compared to the frequencies of the excited modes in resonator $A$, then the SQUID remains adiabatically in its ground state. In this case, we can neglect the first term in the boundary condition Eq.~(\ref{eq:bc_res_A_squid}), and write
\begin{eqnarray}
\label{eq:bc_res_A_squid_diff}
\Phi_A(d, t) + \Delta d(\Phi_{\rm ext}) \partial_x \Phi_A(0, t) = 0,
\end{eqnarray}
where
\begin{eqnarray}
\label{eq:delta_d_eff}
\Delta d(\Phi_{\rm ext}) = \left(\frac{\Phi_0}{2\pi}\right)^2  \frac{1}{L_0E_J(\Phi_{\rm ext})}.
\end{eqnarray}
This can be interpreted as an effective length that can be tuned by the externally applied magnetic
flux $\Phi_{\rm ext}$. If this effective length is small compared to the length scale at which $\Phi_A(x)$
varies substantially, i.e., small compared to the wavelength, then Eq.~(\ref{eq:bc_res_A_squid_diff}) can
be viewed as a differential. If we imagine that the transmission line uniformly extends beyond the
point $x=d_A$ for an additional length $\Delta d(\Phi_{\rm ext})$, we can then rewrite the boundary condition Eq.~(\ref{eq:bc_res_A_squid_diff}) on the simple form
\begin{eqnarray}
\Phi(d_{\rm eff}(\Phi_{\rm ext}), t) = 0,
\end{eqnarray}
where we have introduced the new effective tunable length of resonator A
\begin{eqnarray}
\label{eq:d_eff}
d_{\rm eff}(\Phi_{\rm ext}) = d_A + \Delta d(\Phi_{\rm ext}).
\end{eqnarray}

\subsubsection{Resonator B}\label{sec:bc-res-B}

In the limit $C_B\rightarrow0$, resonator $B$ is an open-ended at both
$x = 0$ and $x = d_B$ (see Fig.~\ref{fig:io-model}), and the corresponding
boundary conditions are therefore
\begin{eqnarray}
\label{eq:bc_res_B_left}
\partial_x \Phi_B(0, t) &=& 0, \\
\label{eq:bc_res_B_right}
\partial_x \Phi_B(d_B, t) &=& 0.
\end{eqnarray}

\subsection{SQUID biasing and effective length}\label{sec:squid}

The externally applied magnetic flux, $\Phi_{\rm ext}$, is partly produced by the field of resonator $B$, and partly by a static background flux, $\Phi_{\rm ext}^0$. Here we assume that the physical dimension of SQUID loop is small compared to the typical length scale at which the field in resonator $B$ varies. We can then decompose the externally applied magnetic flux in a
static bias and a small deviation, 
\begin{equation}
\Phi_{\rm ext} = \Phi^0_{\rm ext} + \Delta\Phi_{\rm ext},
\end{equation}
where the small deviation $\Delta\Phi_{\rm ext}$ is a function of the field amplitude at a single point $x_0$ in resonator $B$.
For now we are not concerned with the detailed form of $\Delta\Phi_{\rm ext}$, and we only
require it to be small compared to $\Phi_0$. Under this condition we can expand
the effective Josephson energy of the SQUID, Eq.~(\ref{eq:SQUID_E_J}), as
\begin{equation}
\label{eq:EJ_Phi_Ext_expanded}
E_J(\Phi_{\rm ext})
\approx
E_J^0 - 2E_{J} \frac{\pi}{\Phi_{0}} {\Delta\Phi_{\mathrm{ext}}} \sin{\left(\pi\frac{\Phi^{0}_{\mathrm{ext}}}{\Phi_{0}} \right )}\end{equation}
where
\begin{equation}
E_J^0 = 2 E_J\cos\left(\pi\frac{\Phi^{0}_{\mathrm{ext}}}{\Phi_{0}}\right).
\end{equation}

Using Eq.~(\ref{eq:EJ_Phi_Ext_expanded}) in the expression for the effective
length associated with the SQUID, Eq.~(\ref{eq:delta_d_eff}), we obtain
\begin{equation}
\Delta d(\Phi_{\rm ext}) 
\approx 
\left(\frac{\Phi_0}{2\pi}\right)^2 \frac{1}{L_0E_J^0}
\left(1 + \pi\frac{\Delta\Phi_{\mathrm{ext}}}{\Phi_{0}}\tan{\left ( \pi\frac{\Phi^{0}_{\mathrm{ext}}}{\Phi_{0}} \right )}\right),
\end{equation}
and to simplify the expressions we write
\begin{equation}
\Delta d(\Phi_{\rm ext}) = \Delta d_0(\Phi_{\rm ext}^0) + \delta d(\Phi_{\rm ext}^0)\Delta\Phi_{\mathrm{ext}}
\end{equation}
with
\begin{eqnarray}
\Delta d_0(\Phi_{\rm ext}^0) &=& \left(\frac{\Phi_0}{2\pi}\right)^2 \frac{1}{L_0E_J^0}, \\
\delta d(\Phi_{\rm ext}^0) 
&=& 
\frac{1}{2}\left(\frac{\Phi_0}{2\pi}\right) \frac{1}{L_0E_J^0}
\tan{\left (\pi\frac{\Phi^{0}_{\mathrm{ext}}}{\Phi_{0}} \right )},
\end{eqnarray}
and
\begin{eqnarray}
\label{eq:d_eff_ext}
d_{\rm eff}(\Phi_{\rm ext}) 
&=& d_{\rm eff}^0(\Phi_{\rm ext}^0) + \delta d(\Phi_{\rm ext}^0)\Delta\Phi_{\mathrm{ext}}, \\
d_{\rm eff}^0(\Phi_{\rm ext}^0)
&=& d_A + \Delta d_0(\Phi_{\rm ext}^0).
\end{eqnarray}

%-------------------------------------------------------------------------------
% Modes
%
\subsection{Fields and modes}\label{sec:modes}

Given the quantum description of the flux field in the two resonators given
in Eq.~(\ref{eq:quantum_field}), we are now interested in using the boundary
conditions given in the previous section to derive the adiabatic modes
for the two resonators.

\subsubsection{Resonator A}\label{sec:modes-res-A}

With the two boundary conditions Eqs.~(\ref{eq:bc_res_A_left},\ref{eq:bc_res_A_squid}),
corresponding to an open and a short circuit, respectively, resonator $A$ becomes a
$\lambda/4$ resonator. In particular, imposing the two boundary conditions results in
the constraint $\cos(k^A_\omega d_{\rm eff}(\Phi_{\rm ext})) = 0$, which is satisfied
with the frequencies $\omega^A_n = \frac{\pi}{2}(2n+1)v/d_{\rm eff}(\Phi_{\rm ext})$. The
field, written in terms of the corresponding mode functions, becomes
\begin{eqnarray}
\label{eq:field-mode-expansion-res-A}
\Phi_A(x,t) &=& \sqrt{\frac{Z_0\hbar}{2\pi}}\sum_n \sqrt{\frac{\omega_{d_A}}{\omega^A_n}}
\cos\left(\frac{\pi(2n+1)x}{2d_{\rm eff}(\Phi_{\rm ext})}\right) \nonumber\\
&&\times (a_n e^{-i\omega^A_n t} + \text{h.c.}),
\end{eqnarray}
where $\omega_{d_A} = 2\pi v/d_{\rm eff}(\Phi_{\rm ext})$ is the full-wavelength frequency of the resonator of length $d_{\rm eff}(\Phi_{\rm ext})$,
and $a_n$ is the annihilation operator of the $n$th mode, which satisfies $[a_n, a_m^\dagger] = \delta_{nm}$.
Here the field is written in terms of the instantaneous, or adiabatic, mode functions
for resonator A, for a given applied magnetic flux $\Phi_{\rm ext}$.

\subsubsection{Resonator B}\label{sec:modes-res-B}

With the two boundary conditions Eqs.~(\ref{eq:bc_res_B_left}, \ref{eq:bc_res_B_right}),
which both are open-ended terminations, resonator $B$ becomes a $\lambda/2$ resonator.
In particular, imposing these two boundary conditions results in the constraint 
$\sin(k^B_\omega d_B) = 0$, which is satisfied with the frequencies 
$\omega^B_n = \pi nv/d_B$. Writing the field in terms of the corresponding
mode functions yields 
\begin{eqnarray}
\label{eq:field-mode-expansion-res-B}
\Phi_B(x,t) &=& \sqrt{\frac{Z_0\hbar}{2\pi}}\sum_n \sqrt{\frac{\omega_{d_B}}{\omega^B_n}}
\cos\left(\frac{\pi nx}{d_B}\right) \nonumber\\
&& \times (b_n e^{-i\omega^B_n t} + \text{h.c.}),
\end{eqnarray}
where $\omega_{d_B} = 2\pi v/d_B$, and $b_n$ is the annihilation operator of the $n$th mode, satisfying $[b_n, b_m^\dagger] = \delta_{nm}$.

%-------------------------------------------------------------------------------
% Effective Hamiltonian
%
\section{Effective Hamiltonian}\label{sec:effective-hamiltonian}

Using the adiabatic modes derived in the previous section, and their corresponding annihilation operators, 
we can write the Hamiltonian for the two resonators on the form
\begin{equation}
\label{eq:hamiltonian_adiabatic}
H =
\sum_n \hbar \omega^A_n a_n^\dagger a_n +
\sum_n \hbar \omega^B_n b_n^\dagger b_n.
\end{equation}

Assuming that $\delta d(\Phi_{\rm ext}^0)\Delta\Phi_{\rm ext} \ll d_{\rm eff}^0(\Phi_{\rm ext})$, we can now
use Eq.~(\ref{eq:d_eff_ext}) to write the mode frequency for resonator $A$ as
\begin{eqnarray}
\omega^A_n &=& \frac{\pi}{2}\frac{(2n+1)v}{d_{\rm eff}^0(\Phi_{\rm ext}^0) + \delta d(\Phi_{\rm ext}^0)\Delta\Phi_{\rm ext}} \nonumber\\
&\approx& \tilde{\omega}_n^A \left(1 - \frac{\delta d(\Phi_{\rm ext}^0)}{d_{\rm eff}^0(\Phi_{\rm ext}^0)}\Delta\Phi_{\rm ext}\right),
\end{eqnarray}
where $\tilde{\omega}_n^A = \frac{\pi}{2}(2n+1)v / d_{\rm eff}^0(\Phi_{\rm ext}^0)$.
Inserting this expression in the Hamiltonian Eq.~(\ref{eq:hamiltonian_adiabatic}), 
we obtain an effective Hamiltonian
\begin{eqnarray}
H &=&
\sum_n \hbar \tilde{\omega}^A_n a_n^\dagger a_n +
\sum_n \hbar \omega^B_n b_n^\dagger b_n \nonumber\\
&-& 
\sum_n \hbar \omega^A_n\frac{\delta d(\Phi_{\rm ext}^0)}{d_{\rm eff}^0(\Phi_{\rm ext}^0)} \Delta\Phi_{\mathrm{ext}} a_n^\dagger a_n.
\end{eqnarray}
This Hamiltonian is valid under the approximation that the modes of  resonator $A$
instantaneously adjust to changes in the applied magnetic flux $\Delta\Phi_{\rm ext}$,
which are due to the dynamics of the field in resonator $B$. This means that we require
$\omega_A \gg \omega_B$.

We now assume that the deviation of the external bias flux from the 
static bias $\Phi_{\rm ext}^0$ takes the form
\begin{equation}
\label{eq:Delta_Phi_ext}
\Delta\Phi_{\mathrm{ext}} = \Phi_0 \sum_n G_n (b_n + b_n^\dagger),
\end{equation}
where $G_n$ is the effective coupling strength between the $n$th mode and the SQUID,
including for example geometric factors, and the normalized mode amplitude at
the point of the SQUID. This form will be motivated later when explicit coupling
geometries are considered.
With this form of $\Delta\Phi_{\mathrm{ext}}$, the effective Hamiltonian takes the
form
\begin{eqnarray}
H &=& 
\sum_n \hbar \tilde{\omega}^A_n a_n^\dagger a_n +
\sum_n \hbar \omega^B_n b_n^\dagger b_n \nonumber\\
&-& \sum_n \hbar \tilde{\omega}^A_n\frac{\delta d(\Phi_{\rm ext}^0)}{d_{\rm eff}^0(\Phi_{\rm ext}^0)} 
\Phi_0 \sum_m G_m (b_m + b_m^\dagger)a_n^\dagger a_n. \nonumber\\
\end{eqnarray}

\subsection{Optomechanical Hamiltonian}\label{sec:optomech-hamiltonian}

If we restrict the dynamics of the system to only involve the two
fundamental modes (i.e., by not exciting any higher modes), we obtain a 
simplified two-mode Hamiltonian
\begin{equation}
\label{eq:hamiltonian-optomech}
H = \hbar \omega_A a^\dagger a 
  + \hbar \omega_B b^\dagger b 
  - \hbar g_0 a^\dagger a (b + b^\dagger),
\end{equation}
where, for brevity, we have dropped the indices on the annihilation operators
and the mode frequencies. Here
\begin{eqnarray}
\label{eq:coupling-strength}
g_0 =  \omega_A F(\Phi^0_{\rm ext}) G_1, 
\end{eqnarray}
is the coupling strength between the two resonators, and
\begin{eqnarray}
\label{eq:F-ext}
F(\Phi^0_{\rm ext}) &=& \Phi_0 \frac{\delta d(\Phi_{\rm ext}^0)}{d_{\rm eff}^0(\Phi_{\rm ext}^0)}.
\end{eqnarray}
The coupling strength is comprised of two factors, in addition to the frequency factor $\omega_A$: 
($i$) A factor $F(\Phi^0_{\rm ext})$ that depends on the properties and the bias conditions of the SQUID,
and ($ii$) a factor $G_1$ that depends of the geometric arrangement of the SQUID and the resonators.
To produce a large coupling strength we are interested in maximizing both of these factors.

The Hamiltonian Eq.~(\ref{eq:hamiltonian-optomech}) is on the standard optomechanical form, and the device we consider here
is therefore analogous with an optomechanical system. However, in contrast to an optomechanical system, here
both resonators are electrical and the fundamental nonlinear interaction strength $g_0$ can be tuned by
changing the flux bias $\Phi_{\rm ext}^0$. 

As in the optomechanical case \cite{meystre:2013}, we have in the derivation of Hamiltonian Eq.~(\ref{eq:hamiltonian-optomech}) assumed that $\omega_A \gg \omega_B$, so that the field in resonator $A$ adiabatically adjust to the parametrically changing resonance frequency due to the dynamics of resonator $B$. We can compensate for the difference in frequencies by applying a driving field on resonator $A$, with frequency $\omega_d$ and amplitude $\epsilon_A$,  
\begin{eqnarray}
	\label{eq:hamiltonian_optomech_multimode}
	H &=& \hbar\omega_A a^\dagger a + \hbar\omega_B b^\dagger b - g_0 a^\dagger a (b^\dagger + b) \nonumber\\
	&+& (\epsilon_A a e^{-i\omega_d t} + \epsilon_A^* a^\dagger e^{i\omega_d t}),
\end{eqnarray}
and applying the unitary transformation $U = \exp\left[i\omega_d a^\dagger a t\right]$, which makes the drive terms time-independent,
\begin{eqnarray}
	\label{eq:hamiltonian_optomech_multimode}
	H &=& \hbar\Delta_A a^\dagger a + \hbar\omega_B b^\dagger b - g_0 a^\dagger a (b^\dagger + b) \nonumber\\
	&+& (\epsilon_A a + \epsilon_A^* a^\dagger).
\end{eqnarray}
Here $\Delta_A = \omega_A - \omega_d$, and if we chose $\Delta_A = \omega_B$, i.e.,
$\omega_d = \omega_A - \omega_B$, the two resonators are effectively resonant. Furthermore, if the amplitude
of the applied driving field $\epsilon_A$ is large, we can linearize the coupling by applying
the unitary displacement transformation $D(\alpha) = \exp\left[\alpha a^\dagger - \alpha^*a\right]$, 
where $\alpha = \epsilon_A/\Delta_A$, and neglecting the term $\hbar g_0 a^\dagger a (b^\dagger + b)$. 
The linearized Hamiltonian is
\begin{eqnarray}
H &=& \hbar\Delta_A a^\dagger a 
  + \hbar\omega_m b^\dagger b 
  %- \hbar g_0 a^\dagger a (b + b^\dagger) 
\nonumber\\
  &+& \hbar g_0 \alpha (a + a^\dagger) (b + b^\dagger)
  - \hbar g_0 |\alpha|^2 (b + b^\dagger),
\end{eqnarray}
and here the strength of the effective linear coupling, $g_0 \alpha$, is proportional to the driving amplitude.
This is commonly used in optomechanics to enhance the coupling strength when the fundamental 
coupling strength $g_0$ itself is too small. This linear coupling regime has several
important applications \cite{meystre:2013,aspelmeyer:2013}, including state transfer, sideband cooling, and 
parametric amplification. Also, in hybrid electro-optomechanical systems, it has been shown that strong
Kerr-nonlinearities can be realized in this weak coupling regime \cite{lu:2013}.

If, on the other hand, the fundamental coupling strength $g_0$ is comparable
to $\omega_B$, it is instructive to apply the unitary transformation 
$U = \exp\left[-g_0a^\dagger a(b^\dagger - b)/\omega_B\right]$, after which the Hamiltonian
Eq.~(\ref{eq:hamiltonian-optomech}) takes the form 
\begin{equation}
H = \hbar \Delta_A a^\dagger a + \hbar \omega_B b^\dagger b + \hbar \frac{g_0^2}{\omega_B} (a^\dagger a)^2.
\end{equation}
This Hamiltonian includes a nonlinear Kerr term, i.e., an effective photon-photon interaction term, with coupling strength $g^2_0/\omega_B$.
This regime has recently been actively studied theoretically in optomechanics \cite{nunnenkamp:2011,rabl:2011,nunnenkamp:2012},
and it has been shown to feature interesting phenomena, such as photon blockade effects \cite{rabl:2011},
and multiple cooling resonances \cite{nunnenkamp:2012}, and allowing for the generation of nonclassical states \cite{bose:1997,qian:2012,kronwald:2013,nation:2013}.

%In the following section we evaluate the fundamental coupling strength $g_0$ for the device we consider here, and we show that both the weak and strong coupling regimes discussed above should be obtainable.

%-------------------------------------------------------------------------------
% Coupling strength
%
\section{Coupling strength}\label{sec:coupling-strength}

\begin{figure}[t]
\begin{center}
\includegraphics[width=8.5cm]{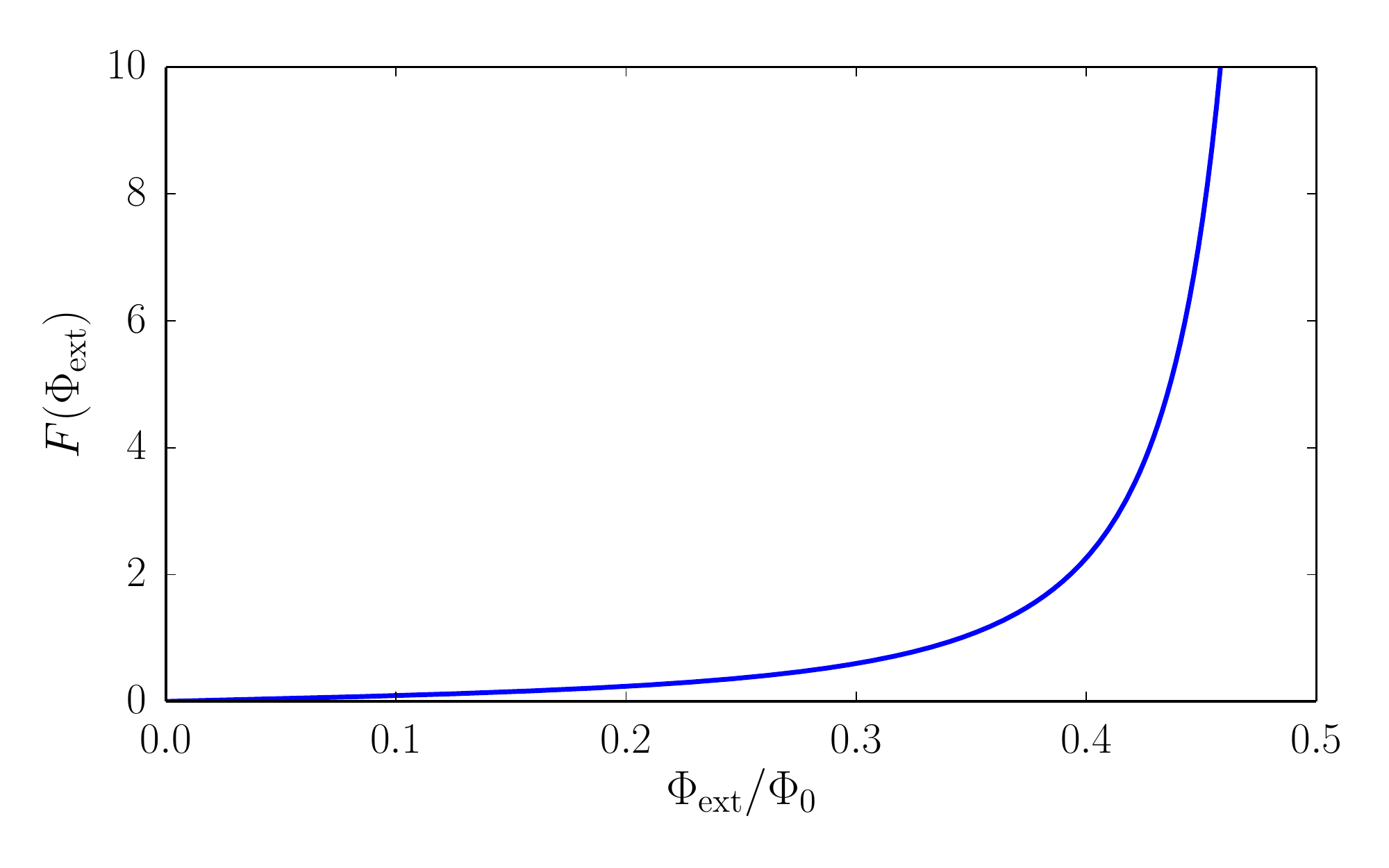}
\caption{(color online) The $\Phi_{\rm ext}$-dependence of the factor 
$F(\Phi_{\rm ext})$ that appears in the coupling strength [see Eq.~(\ref{eq:coupling-strength})].}
\label{fig:F_ext-vs-Phi_ext}
\end{center}
\end{figure}

In this section, we explicitly evaluate the coupling strength $g_0$ for two possible coupling
geometries. We first turn our attention to the factor $F(\Phi_{\rm ext})$, which only
depends on the properties and the bias conditions of the SQUID. The explicit
form of $F(\Phi_{\rm ext})$ is
\begin{eqnarray}
F(\Phi_{\rm ext}) =
\pi
\frac{\Delta d_0(\Phi_{\rm ext})}{d_A + \Delta d_0(\Phi_{\rm ext})}
\tan\left(\pi\frac{\Phi_{\mathrm{ext}}}{\Phi_{0}}\right),
\end{eqnarray}
which is shown graphically in Fig.~(\ref{fig:F_ext-vs-Phi_ext}).
It is clear that $F(\Phi_{\rm ext})$ can be made small by tuning $\Phi_{\rm ext}$ to 0, and
also that it can be tuned to the order of unity, or even much
larger, by letting $\Phi_{\rm ext}$ approach $\frac{1}{2}\Phi_0$. However, when
$\Phi_{\rm ext}$ approach $\frac{1}{2}\Phi_0$, the plasma frequency of the
SQUID decrease rapidly, and an assumption in deriving the effective
Hamiltonian was that this plasma frequency must be much larger than the resonance
frequency $\omega_A$. This prohibits tuning $\Phi_{\rm ext}$ too close to
$\frac{1}{2}\Phi_0$. However, with reasonable values of $\Phi_{\rm ext}/\Phi_0$
it is realistic to obtain $F(\Phi_{\rm ext})$ of the order of 1
(e.g., $\Phi_{\rm ext}/\Phi_0 \approx 0.3 \sim 0.4$). With an increased
plasma frequency of the SQUID, i.e., a large Josephson energy $E_J$, $\Phi_{\rm ext}$
could possibly be further increased.

\begin{figure}[t]
\begin{center}
\includegraphics[width=8.5cm]{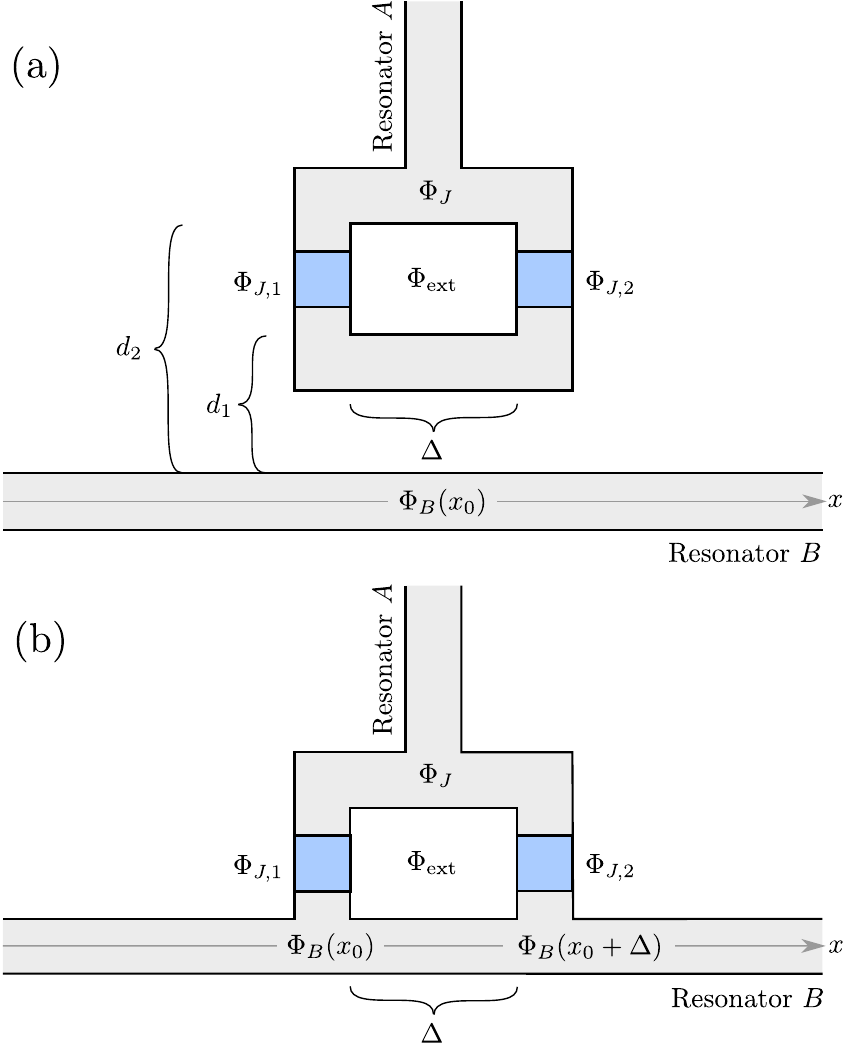}
\caption{(color online) Two possible coupling geometries: (a) An inductive
coupling design, where the magnetic field from resonator $B$ couples inductively
to the SQUID loop. (b) An alternative coupling design, with potentially larger
coupling strength, where the SQUID loop is galvanically connected to resonator $B$. }
\label{fig:coupling-model}
\end{center}
\end{figure}

In addition to the factor $F(\Phi_{\rm ext})$, the coupling strength Eq.~(\ref{eq:coupling-strength})
also contains the factor $G_1$, which depends on the detailed geometry of the coupling. Below
we estimate the numerical values of $G_1$ for two possible geometries shown in 
Fig.~\ref{fig:coupling-model}.

\subsection{Inductive coupling}\label{sec:couping-strength-ind}

A schematic illustration of a coupling design where the magnetic field of resonator
$B$ couples inductively to the SQUID loop is shown in Fig.~\ref{fig:coupling-model}(a).
An exact calculation of the coupling strength for this design would
require detailed modeling of how the magnetic field extends around the microstrips
that define the microwave resonators. However, an estimate of the coupling strength
can be obtained by assuming that the magnetic field $B(x, r)$ takes the form of that surrounding
a perfect line conductor. In this case
\begin{equation}
B(x, r) = \frac{\mu_0 I_B(x)}{2\pi r},
\end{equation}
where $r$ is the radial distance from the conductor, $I_B(x)$ is the current 
at position $x$, and $\mu_0$ is the permeability of free space. The
current $I_B(x)$ can be evaluated using the expression for the field in terms of
the mode functions, Eq.~(\ref{eq:field-mode-expansion-res-B}),
\begin{eqnarray}
\label{eq:current-B}
I_B(x_0) &=&
-L_0^{-1} \partial_x \Phi_B(x_0) \nonumber\\
&=&
L_0^{-1}\sqrt{\frac{Z_0\hbar}{2\pi}}\sum_n \sqrt{\frac{\omega_{d_B}}{\omega^B_n}}
\frac{\pi n}{d_B} \times  \nonumber\\
&& \sin\left(\frac{\pi nx_0}{d_B}\right) (b_n e^{-i\omega_n^B t} + \text{h.c.})
\end{eqnarray}
The strongest coupling strength (for the fundamental mode $n=1$, as well as other odd-$n$ modes) is obtained
by placing resonator $A$ so that it couples to resonator $B$ at the midpoint
$x_0 = d_B/2$, in which case we have
\begin{eqnarray}
\label{eq:current-B-x0}
I_B(x_0) &=&
L_0^{-1}\sqrt{\frac{Z_0\hbar}{2\pi}}\frac{1}{d_B} \sum_n  \sqrt{\frac{\omega_{d_B}}{\omega^B_n}} \pi n
 \times \nonumber\\
&& \sin\left(\frac{\pi n}{2}\right) (b_n e^{-i\omega_n^B t} + \text{h.c.}).
\end{eqnarray}

The magnetic flux through the SQUID due to the field from the resonator $B$
can be written
\begin{equation}
\Delta\Phi_\mathrm{ext} = \int_S B \cdot \mathrm{d}S = \int_{d_1}^{d_2}\mathrm{d}r \int_{x_0}^{x_0+\Delta} \mathrm{d}x B(x, r),
\end{equation}
and assuming that $I_B(x)$ is constant over $[x_0, x_0+\Delta]$, we obtain
\begin{equation}
\Delta\Phi_\mathrm{ext}
= \Delta \frac{\mu_0 I(x_0)}{2\pi} \int_{d_1}^{d_2} \frac{\mathrm{d}r}{r}
= \Delta \frac{\mu_0 I(x_0)}{2\pi} \log(d_2/d_1).
\end{equation}
Using the expression for the current given in Eq.~(\ref{eq:current-B}), we have
\begin{eqnarray}
\Delta\Phi_\mathrm{ext}
&=& 
\frac{\mu_0 \Delta \log(d_2/d_1)}{2 L_0 d_B} 
\sqrt{\frac{Z_0\hbar}{2\pi}}\sum_n \sqrt{\frac{\omega_{d_B}}{\omega^B_n}}n
\times \nonumber \\
&& \sin\left(\frac{\pi n}{2}\right) (b_n e^{-i\omega^B_n t} + \text{h.c.}).
\end{eqnarray}
This can be written on the form of Eq.~(\ref{eq:Delta_Phi_ext}) with
\begin{eqnarray}
G^{\rm ind}_n &=& 
\frac{\mu_0 \Delta n \log(d_2/d_1)}{2L_0 \Phi_0 d_B} 
\sqrt{\frac{Z_0\hbar}{2\pi}\frac{\omega_{d_B}}{\omega^B_n}} \sin\left(\frac{\pi n}{2}\right),\;\; 
\end{eqnarray}
and, in particular, for the fundamental mode ($n=1$), which we are most interested in, we have
\begin{eqnarray}
G^{\rm ind}_1 = 
\mu_0L_0^{-1} \log(d_2/d_1) \frac{\Delta}{2d_B} \frac{1}{\Phi_0} \sqrt{\frac{Z_0\hbar}{2\pi}}  \sqrt{\frac{\omega_{d_B}}{\omega^B_1}}.
\end{eqnarray}

\subsection{Galvanic coupling}\label{sec:couping-strength-galv}

An alternative coupling design, which could produce stronger coupling,
is shown in Fig.~\ref{fig:coupling-model}(b). In this case, part of the SQUID
loop is galvanically connected to resonator $B$, and the fluxoid quantization
condition for the SQUID loop is
\begin{equation}
\label{eq:fluxoid_quantization_condition}
\Phi_{\rm ext} = \Phi_{J,1} - \Phi_{J,2} + \Phi_B(x_0+\Delta) - \Phi_B(x_0).
\end{equation}
Assuming that the field in the resonator $B$ varies only slightly between $x_0$ and $x_0 + \Delta$,
we can linearize and write the difference $\Phi_B(x_0+\Delta) - \Phi_B(x_0)$ in
Eq.~(\ref{eq:fluxoid_quantization_condition}) as a differential
\begin{equation}
\Phi_{\rm ext} = \Phi_{J,1} - \Phi_{J,2} + \Delta \partial_x \Phi_B(x_0).
\end{equation}
We can now use this constraint to proceed as usual and eliminate one phase variable,
and introduce the new variable $\Phi_J$ for the remaining SQUID degree of freedom
\begin{eqnarray}
\Phi_{J,1} &=& \Phi_J + \frac{1}{2}\left(\Phi_{\rm ext} - \Delta \partial_x \Phi_B(x_0)\right), \\
\Phi_{J,2} &=& \Phi_J - \frac{1}{2}\left(\Phi_{\rm ext} - \Delta \partial_x \Phi_B(x_0)\right).
\end{eqnarray}

Here we identify $\Delta\Phi_\mathrm{ext} = - \Delta \partial_x \Phi_B(x_0)$, and 
using the expression for the field of resonator $B$, Eq.~(\ref{eq:field-mode-expansion-res-B}), we obtain
\begin{eqnarray}
\Delta\Phi_\mathrm{ext} &=& 
\frac{\Delta}{d_B}
\sqrt{\frac{Z_0\hbar}{2\pi}}\sum_n \sqrt{\frac{\omega_{d_B}}{\omega^B_n}} \pi n \times \nonumber\\ && \sin\left(\frac{\pi nx_0}{d_B}\right) 
(b_n e^{-i\omega^B_n t} + \text{h.c.}),
\end{eqnarray}
which can be written on the form of Eq.~(\ref{eq:Delta_Phi_ext}) with
\begin{equation}
G^{\rm galv}_n = 
\pi n \frac{\Delta}{d_B}
\frac{1}{\Phi_0}
\sqrt{\frac{Z_0\hbar}{2\pi}}
\sqrt{\frac{\omega_{d_B}}{\omega^B_n}}
\sin\left(\frac{\pi n x_0}{d_B}\right).
\end{equation}
Again, we are most interested in the coupling strength for the fundamental mode ($n=1$),
and for $x_0 = d_B/2$, we have
\begin{equation}
G^{\rm galv}_1 = \pi\frac{\Delta}{d_B} \frac{1}{\Phi_0}\sqrt{\frac{Z_0\hbar}{2\pi}} \sqrt{\frac{\omega_{d_B}}{\omega^B_1}}.
\end{equation}

\subsection{Total coupling strength}\label{sec:couping-strength-total}

\begin{figure}[t]
\begin{center}
\includegraphics[width=8.5cm]{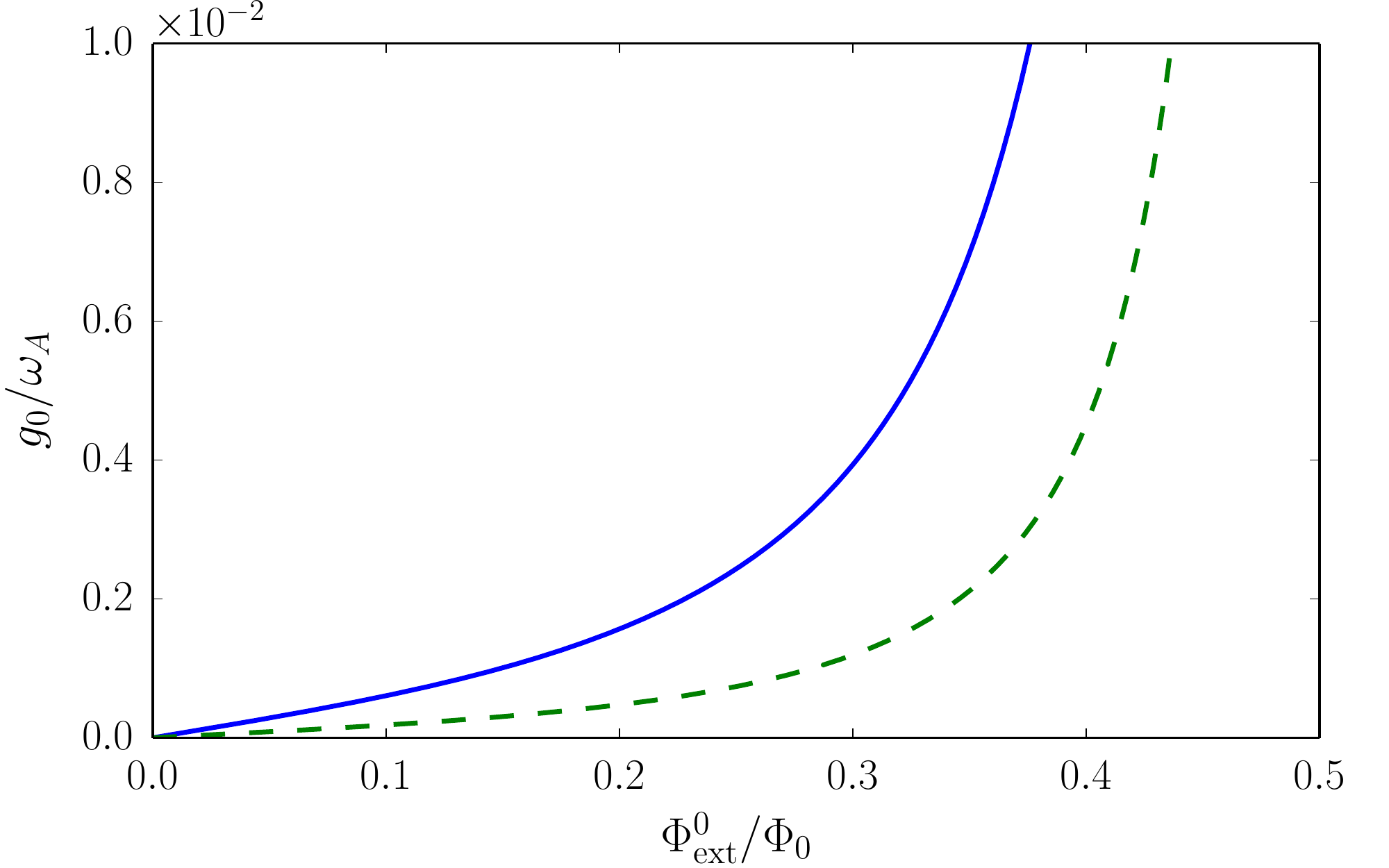}
\caption{(color online) The normalized coupling strength $g_0/\omega_A$ as
a function of the external flux bias $\Phi_{\rm ext}$, for the galvanic (blue solid)
and inductive (green dashed) coupling designs. The parameters used
to evaluate Eq.~(\ref{eq:coupling-strength-final}) were: $Z_0 \approx 50\;\Omega$,
$\omega_A = 10$ GHz, $\omega_B = 1$ GHz, $d_A = d_B/20 = 3$ mm, 
$L_0 = 4.57\cdot10^{-7}$ H/m, $C_0 = 1.46\cdot10^{-10}$ H/m, $\Delta/d_B = 10$\%,
and $E_J = 4.11\cdot10^{-22}$ J.}
\label{fig:coupling-strength-vs-bias}
\end{center}
\end{figure}
\begin{figure}[t]
\begin{center}
\includegraphics[width=8.5cm]{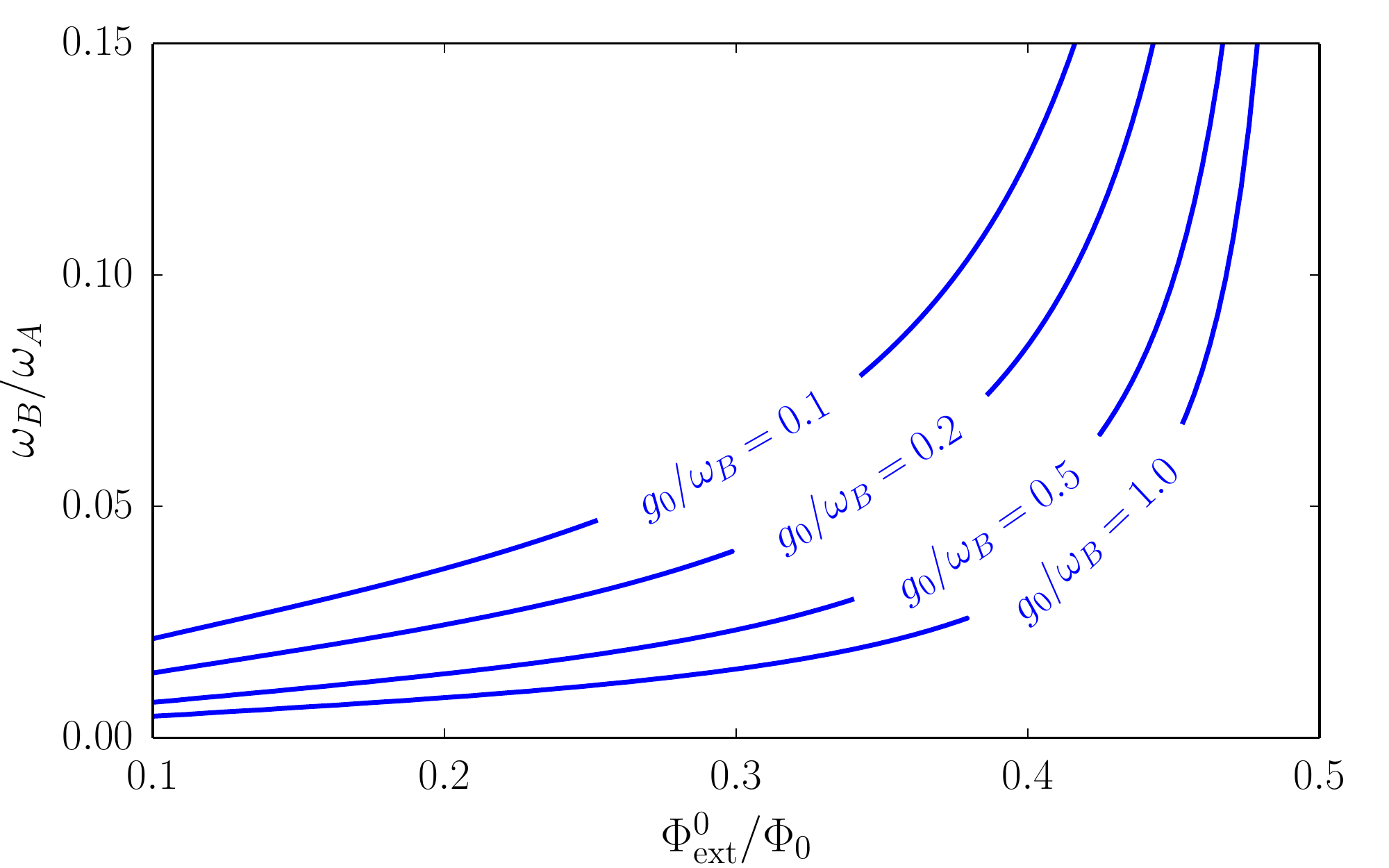}
\caption{(color online) Contours of $g_0/\omega_B$, the ratio of the fundamental coupling
strength to the frequency of resonator $B$, as a function of the ratio between the frequencies
of resonator $B$ and $A$, $\omega_B/\omega_A$, and the external flux bias $\Phi^0_{\rm ext}/\Phi_0$, for the 
galvanic coupling design.
Here we used the same parameters as in Fig.~\ref{fig:coupling-strength-vs-bias}, except that here
$\omega_A$ is varied.
}
\label{fig:coupling-strength-frequency-ratio-and-squid-size}
\end{center}
\end{figure}

The ratio of the coupling strengths obtained for the inductive and galvanic coupling
designs, assuming equal $\Delta$, is
\begin{eqnarray}
\frac{G_1^{\rm galv}}{G_1^{\rm ind}} 
= \frac{2\pi}{\mu_0L_0^{-1}\log(d_2/d_1)}.
\end{eqnarray}
With $d_2 = 2 d_1$, $\mu_0 \approx 1.26\cdot10^{-6}$ H/m, and $L_0 = 4.57\cdot10^{-7}$ H/m, this ratio is approximately
$3.3$, demonstrating that galvanic coupling design is slightly more efficient.

The explicit form of the total coupling strength $g_0$, using the more favorable galvanic coupling design, is
\begin{eqnarray}
\label{eq:coupling-strength-final}
g_0 
=
\omega^A_1
\frac{\Delta}{d_B}\frac{\Delta d_0(\Phi_{\rm ext}^0)}{d_A + \Delta d_0(\Phi_{\rm ext}^0)}
\frac{\pi^2}{\Phi_0}\sqrt{\frac{Z_0\hbar}{2\pi}\frac{\omega_{d_B}}{\omega^B_1}}
\tan\left( \pi\frac{\Phi^{0}_{\mathrm{ext}}}{\Phi_{0}}\right). \nonumber\\
\end{eqnarray}
This expression is shown graphically in Fig.~\ref{fig:coupling-strength-vs-bias}
and Fig.~\ref{fig:coupling-strength-frequency-ratio-and-squid-size},
for the specific parameters given in the captions.
It is clear that $g_0 \ll \omega_A$, as expected and required, but it is not necessary that 
$g_0 \ll \omega_B$, since $\omega_B$ should be at least an order of magnitude smaller than
$\omega_A$. Also, with resonators with sufficiently large $Q$-factors ($\sim 10^3$),
it should be possible to enter the strong coupling regimes,
where the frequency shift of resonator $A$ due to the presence of a single photon
in resonator $B$ exceeds the linewidth of resonator $A$, i.e., $g_0 > \kappa_A$,
or when a single photon in resonator $A$ displaces resonator $B$ an amount that
exceeds its zero-point fluctuations, i.e., $g_0 > \kappa_B$ and $g_0 \sim \omega_B$.
Here $\kappa_A$ and $\kappa_B$ denotes the relaxation rates of resonator $A$ and $B$, respectively. 

In particular, the single-photon strong-coupling regime \cite{aspelmeyer:2013}, where $g_0 \sim \omega_B$,
should be realizable in the circuit considered here. Figure \ref{fig:coupling-strength-frequency-ratio-and-squid-size} shows
the ratio $g_0/\omega_B$ as a function of the resonator frequencies and the externally applied flux bias.
When the frequency ratio $\omega_B/\omega_A$ is sufficiently small, it should be possible to reach
$g_0 \sim \omega_B$ for reasonable values of $\Phi^0_{\rm ext}/\Phi_0$ (i.e., not too close to 0.5).
The device we consider here is therefore a possible candidate for realizing 
an analog of an optomechanical system in this strong-coupling regime.

%-------------------------------------------------------------------------------
% Arrays of coupled resonators
%
\section{Arrays of coupled resonators}\label{sec:arrays}

Using the nonlinear coupling mechanism for microwave resonators that we have 
investigated here, it is straightforward to imagine all-electrical networks,
or arrays, of analog optomechanical resonators.
Superstructures of optomechanical systems, for example optomechanical crystal
arrays \cite{chang:2011}, have recently received considerable attention
\cite{xuereb:2012} for their potential applications
in quantum information processing \cite{schmidt:2012} and quantum simulation \cite{tomadin:2012,ludwig:2013}.
% for example
%by many-body physics applications, including phase-transitions between localized
%and propagating excitations
%
Implementing
such systems with the all-electrical SQUID-coupled resonators considered here 
could have advantages in terms of designability, coupling strengths and
in-situ controllability. Also, since all resonators in this architecture are electrical,
they could all be probed and driven using the same microwave technologies.
It is also relatively easy to construct various topologies
among the coupled resonators, as we show below.

\begin{figure}[t]
\begin{center}
(a)\includegraphics[width=8.5cm]{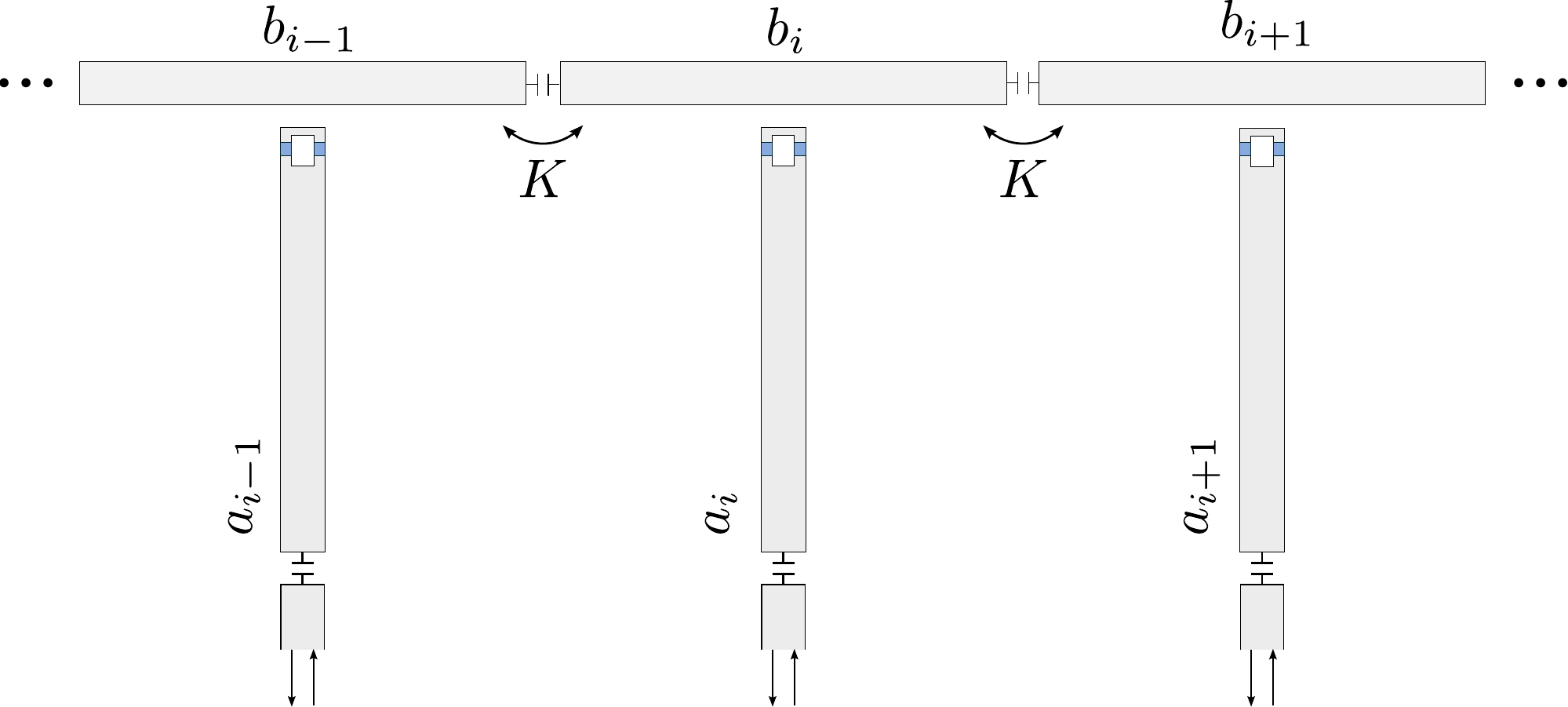}\vspace{1cm}
(b)\includegraphics[width=8.5cm]{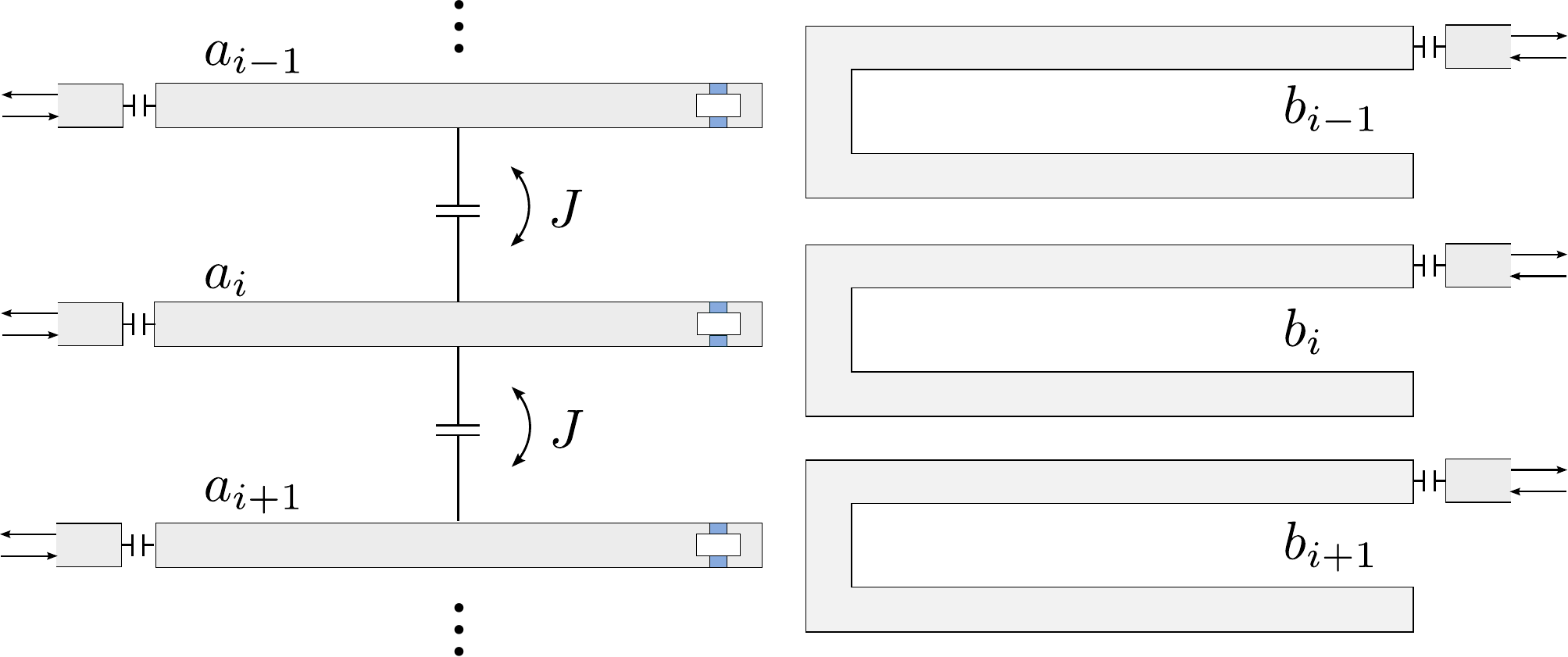}\vspace{1cm}
(c)\includegraphics[width=8.5cm]{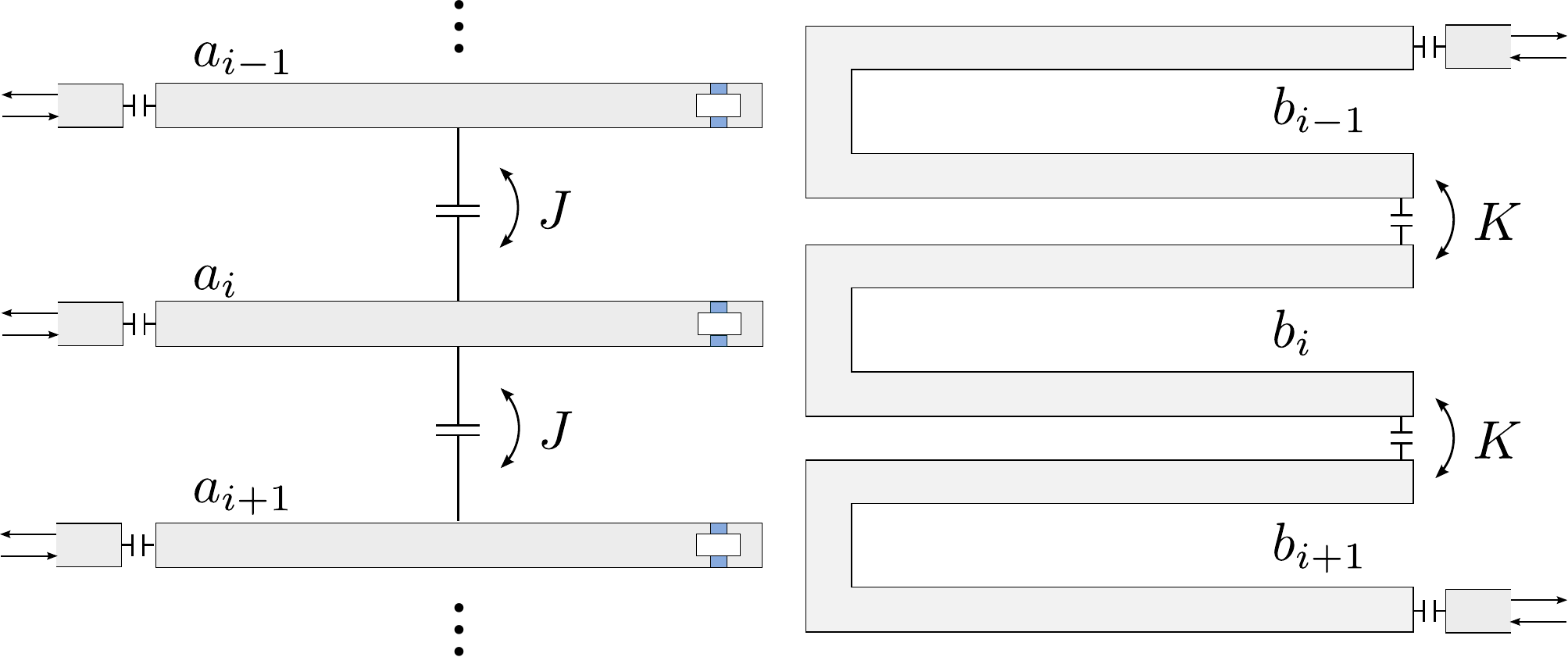}
\caption{(color online) Three possible networks of nonlinearly coupled
microwave resonators. These circuits represents all-electrical networks of 
resonators that are analogous to arrays of optomechanical systems.
The electrical implementation here replaces the mechanical resonator with an electrical
resonator, while keeping the nonlinear interaction. With different layouts
it is possible to create networks where the ``mechanical'' (a), ``optical'' (b)
or both (c) systems are coupled.}
\label{fig:resonator-array}
\end{center}
\end{figure}

In a network of linearly coupled optomechanical systems, we can write the Hamiltonian for a single
unit consisting of two nonlinearly coupled resonators as
\begin{eqnarray}
H_i &=& \omega_A^{(i)} a_i^\dagger a_i + \omega_B^{(i)} b_i^\dagger b_i
- g_ia_i^\dagger a_i (b_i + b_i^\dagger) \nonumber\\
&+&
\mathcal{\epsilon}_A^{(i)}(a_i + a_i^\dagger) +
\mathcal{\epsilon}_B^{(i)}(b_i + b_i^\dagger),
\end{eqnarray}
where we have included driving fields applied to the two resonators,
with amplitudes $\mathcal{\epsilon}_A^{(i)}$ and $\mathcal{\epsilon}_B^{(i)}$,
for resonator $A$ and $B$, respectively.
Apart from the additional driving fields, this Hamiltonian has the form of the
effective Hamiltonian Eq.~(\ref{eq:hamiltonian-optomech}). The driving fields
can be easily applied via the capacitive coupling to the
external transmission lines.

The Hamiltonian of a general linearly-coupled nearest-neighbor array of these
unit systems can then be written on the form
\begin{eqnarray}
\label{eq:hamiltonian-array}
H = \sum_iH_i 
&+& J \sum_{\left<i,j\right>}\left(a_i^\dagger a_j + a_i a_j^\dagger\right) \nonumber\\
&+& K \sum_{\left<i,j\right>}\left(b_i^\dagger b_j + b_i b_j^\dagger\right).
\end{eqnarray}
Here $K$ and $J$ are the strengths of the linear coupling between resonators in different unit cells.
In the case of electrical resonators, this type of coupling is realized using capacitive coupling between
the resonators, and the strength of the coupling can be controlled
in the design of the corresponding capacitances. Whether either or both of $K$ and $J$ are nonzero
in a particular implementation depends on the layout. In Fig.~\ref{fig:resonator-array}, three
possible layouts are shown schematically. The layout in Fig.~\ref{fig:resonator-array}(a)
is a realization of a system described by the Hamiltonian Eq.~(\ref{eq:hamiltonian-array}) with
$J = 0$ and $K > 0$ (i.e., coupled ``mechanical'' systems), Fig.~\ref{fig:resonator-array}(b)
is a realization of a system with $K = 0$ and $J > 0$ (i.e., coupled ``optical'' systems), 
and Fig.~\ref{fig:resonator-array}(c)
is a realization of a system where both $J > 0$ and $K > 0$ (i.e., both the ``mechanical'' and
the ``optical'' systems are coupled). 

%-------------------------------------------------------------------------------
% Conclusions
%
\section{Conclusions}\label{sec:conclusions}

We have introduced and analyzed a nonlinear coupling mechanism for 
superconducting microwave resonators. With the proposed coupling scheme,
it is possible to realize analogs of optomechanical systems in an all-electrical
circuit. The optomechanical-like interaction can be made both strong and tunable 
through an externally applied flux bias. This all-electrical realization of
optomechanical-like systems could therefore be used to explore the optomechanical model
in new interesting regimes. We have also discussed potential applications of this
circuit realization of the optomechanical model as an alternative way of
implementing arrays of ``optomechanical'' systems, which can be used, for example,
in quantum simulator applications. We believe that the introduced nonlinear coupling
provides new opportunities for implementing analogs of quantum systems
in superconducting circuits.
 
\section*{Acknowledgements}

We wish to thank X.-Y.~Lü, H.~Jing, and Jonas Bylander for helpful discussions.
This work was partly supported by the RIKEN iTHES Project, MURI Center for Dynamic Magneto-Optics, 
JSPS-RFBR No.~12-02-92100, Grant-in-Aid for Scientific Research (S),
MEXT Kakenhi on Quantum Cybernetics, and the JSPS-FIRST program.
GJ acknowledges funding from the Swedish Research Council and the European Research Council.

\bibliography{references}
\end{document}